# Recursive index for assessing value added of individual scientific publications

Eldar Knar*

*Altyngarin National Academy of Education, Astana, Kazakhstan*

*eldarknar@gmail.com*

## Abstract

An aggregated recursive K-index is proposed as a new scientometric indicator of added value and scientific research output of individual publications. This index can be used instead of or in addition to the H-index (J.E. Hirsch. *An index to quantify an individual's scientific research output*, arXiv:physics/0508025).

In particular, it is proposed to switch from a pure strategy for assessing the quality and effectiveness of R&D using the H-index (Hirsch index) to a mixed strategy (in the context of publication activity as a combination of cooperative and noncooperative games) using the K-index on subnational and H-index on international or differentiated levels. In the context of a hybrid strategy of the scientist's payoff functions. This transition is correct and in demand for a number of national scientific systems with limited financial, material, infrastructural and linguistic (in terms of the English language) potential. Scientific systems with highly developed indigenous (autochthonous) characteristics are also needed in some scientific areas.

Some results on the counterproductivity of the Hirsch index for the Kazakh scientific system are presented. In particular, a significant increase in the H-index of Kazakh scientists was noted as the degree of scientific localization (national participation) decreased and as the degree of "foreign content" in science-intensive content increased within the framework of international joint collaboration. The ability of the delocalization of local scientific content in joint publications to increase the H-index of Kazakh scientists was discovered. This effect is interpreted and confirmed by the positive and negative correlation coefficients between the Hirsch indices and the form and degree of participation (role functions) of Kazakh scientists in joint publications.

**Acknowledgments and Declaration:** This article was prepared within the programme Trust Fund Study OR 11465474 *"Scientific foundations of modernization of the education system and* science" (2021-2023, Altynsarin National Academy of Education, Astana, Republic of Kazakhstan).

No AI platforms (LLM, such as ChatGPT) were used in writing the manuscript. This is the principled position of the author.

* https://orcid.org/0000-0002-7490-8375

# 1. Introduction

There are grounds to believe that the scientific community needs to transition from H-indexism to H-indexphobia. The number of articles—whether three or three hundred—is not important; what matters is their effectiveness.

Of course, this does not diminish the outstanding and epochal contribution of Hirsch to scientometrics, the methodology of science, or scientific policy. Hirsch created an indicator that defined the development of publication activity for two decades. In fact, the h-index became a motivator and catalyst for the rapid development of scientific research, including research aimed at achieving high publication activity.

However, this activity has become too intense. The quantity of publications in pursuit of the h-index exceeded all reasonable and unreasonable limits, giving a reduction in added value and rise to "dark matter" in science and "shadow science." Something needs to be done about this.

It is time to return to the roots—to quality rather than quantity. However, in a modern format. This will not make enthusiasts of publishing do it less. However, it will provide opportunities for introverted scientists and those who believe that their scientific contribution can be expressed concisely for official scientometric recognition without the need to generate articles day and night.

In terms of the economy, major and recognized scientific publishing houses will not suffer financially. The value of publications will not diminish the least. The number of candidate articles will still significantly exceed the publication potential. However, at the same time, the scientific significance and added value of individual articles will increase significantly. Additionally, the number of "predatory" journals will rapidly decline.

In this regard, a new scientometric index, the K-index, based solely on the scientific value and added value of individual publications, is proposed. This index can be used instead of or in addition to the H-index.

This K-index is based on the following assumptions, interpreted within the framework of the philosophy of scientometrics and phenomenology of game theory.

There are no logical or factual grounds to assert that the scientific value of publications depends on their quantity, and the value of a scientist is determined by the volume of content they produce—such claims do not hold. Modern Nobel laureates typically have a high Hirsch index. However, this does not imply that the Nobel Prize or any other international recognition is a consequence of a high Hirsch index. Rather, a high Hirsch index is often a result of scientific recognition as an outcome of the Matthew effect (Merton, 1968). Awards are generally given for impressive or groundbreaking results, which are not interpreted through the quantity of publications and, consequently, the index.

For instance, professor Ualbai Umirbaev (Wayne State University, Detroit, USA) received the Moore Prize (The 2007 E. H. Moore Research Article Prize, American Mathematical Society, USA) not for his Hirsch index but for one outstanding research article[†]. The H-index correlates but does not identify the scientific significance of a scientist or their work. In other words, the H-index is correct as a bibliometric indicator but not as a scientometric indicator. Clearly, the contribution to science should be determined not by the quantity of publications but by the valuable quality of individual publications.

Therefore, at the core of Jose Hirsch's H-index (Hirsch, 2005) lies the assumption that the effectiveness and productivity of a scientist are determined by the level of their publication activity. This assumption implies that having as many n-papers as possible is a self-evident condition for successful and productive scientific activity. Without this assumption, all of Hirsch's conclusions leading to the H-index lack meaning and logic.

---

[†] I.P. Shestakov, U.U. Umirbaev, Tame and wild automorphisms of rings of polynomials in three variables. J. Amer. Math. Soc. 17 (2004), p 197-227.

However, Hirsch never explained why, for the "assessment and performance evaluations, for example, for hiring and promoting university faculty, awarding grants, etc." (as Hirsch writes), at least 1-2-3 papers are necessary. Of course, if a scientist has something to communicate, there is no reason to restrict them from publishing as much as they want or need to reflect their results. However, publication activity should not become an indicator or criterion of scientific quality and productivity.

One can express their scientific value quite briefly, as Évariste Galois did in several letters, diaries and notes.. One can express their scientific value through the maximum possible number of publications, as Leonardo Euler did. However, it is hardly reasonable to claim that Euler is more valuable for science than Galois because he has 850 more publications and a higher Hirsch index.

A scientist who can encapsulate the essence of their entire scientific life and results in the form of one monograph or a series of books (especially in the humanities, agriculture, and other fields) cannot afford to do so. This is because, in the eyes of modern scientometrics, they are nothing more than an empty space in the scientific landscape. Therefore, to appear worthy, they have to stretch their results across tens and hundreds of publications.

One might argue that during Galois's time and other great scientists, there was no indexing, SCOPUS, or Web of Science. However, if a new Galois with a supernova quantum algebra, the principles of which he outlined in three publications, were to appear today, what would be his Hirsch index?

Furthermore, Hirsch writes a rather peculiar phrase: "Thus, I claim that two individuals with the same h are comparable in terms of their overall scientific impact, even if their total number of papers or their total number of citations differ significantly." This assertion is the basis of scientometric evaluation by scientists and organizations. This somewhat dubious statement allows for the absolutization and fetishization of the Hirsch index within scientific policy. If one scientist with an H-index of 10 has an incomparable and overwhelmingly greater (or simply different) "overall scientific impact" than another scientist with an H-index of 10 (or higher), then the Hirsch index itself loses any meaning as an evaluation and criterion.

However, the basis for this assertion is entirely unclear. In what way can the overall scientific impact of the creator of modern higher algebra, Évariste Galois, with an index of 4, a few letter articles, and an infinite (including uncited) number of citations be comparable to the overall scientific impact of a beginning PhD with an index of 4, dozens of articles and a few citations in four papers? How can the "overall scientific impact" of the creator of Kurt Gödel's "incompleteness theorem", with an H-index of 8 and 18 articles, be comparable to the "overall scientific impact" of a local "luminary of science", with an index of 8 and 180 publications? How can Grigori Perelman (Hirsch index 18-21) or Ualbai Umirbaev (Hirsch index 15), who solved breakthrough scientific problems, be compared to many hundred and thousands of others "of their kind" in the context of equal Hirsch indices?

The dominance of the Hirsch index as a scientometric indicator creates a strategic equilibrium, similar to Nash's equilibrium, in national scientific systems. Other strategies inevitably lead to the loss of a scientist. Most rules and norms stipulate the Hirsch index as a mandatory attribute of scientific activity and scientific significance. In this context – the more, the better (even if the publication is not cited at all). This situation resembles a very poor fishing bait, hoping that some fish will probabilistically bite it.

In this sense, the Hirsch index in the Kazakhstani scientific system (as in many other countries) is purely an economic category. It determines the level of well-being of a scientific individual and the degree of accessibility of financial resources for research. That is, under the conditions of total domination, the Hirsch index in some scientific systems has become an analog of the Gini index of financial stratification (*Gini index*) in the scientific community.

Therefore, the Hirsch index, as an indicator of scientific significance or usefulness, is a pure strategy of the scientific community due to its dominance. However, this forced pure strategy by the Hirsch index hinders or, at least, does not contribute to the development of national scientific systems. In particular, "Hirschmania" generates not only a decrease in the quality of landslides in publications but also a number of other negative tendencies and trends noted in this work. This concerns the diffusion (dilution) of the dominant role functions of a scientist in publications as the Hirsch index increases within international collaborative research. When scientists switch to collaboration to improve their position in the Hirsch index hierarchy.

Hirschmania is destructive not only in the objective sense but also in the subjective sense. For this reason, some researchers consider it a mild form of "mania of grandiosity" caused by a "mental virus" (Brodie, 2004).

## 2. Delocalization of Local Scientific Content to the Boost H-Index

Let us create a table of Kazakhstani researchers (with Kazakhstani citizenship) with an H-Index exceeding 10 in the field of natural sciences (primarily physics, mathematics, chemistry, and biology) from 2013 to 2022, as of July 18, 2023.

**Table 1.** TOP-100-RK: Natural Sciences: Ranking of Kazakhstani Researchers by H-Index in the field of "Natural Sciences" (based on the SCOPUS database only)

| № | Author | H-index | DOC | CIT | FA/ FWCI1 | LA FWCI 2 | CoA/ FWCI 3 | CorA/ FWCI 4 | SA/ FWCI 5 |
|---|---|---|---|---|---|---|---|---|---|
| 1 | Myrzakulov Ratbay | 48 | 294 | 7 765 | 9% 1.775 | 61% 1.292 | 30% 1.337 | 3% 0.732 | 0 |
| 2 | Zhautykov Bulat | 42 | 157 | 7 117 | 0 | 11% 0.697 | 89% 2.403 | 0% 0 | 0 |
| 3 | Zdorovets Maxim | 41 | 401 | 6 391 | 9% 3.008 | 43% 1.457 | 48% 1.583 | 6% 1.921 | 0 |
| 4 | Kozlovskiy Artem | 40 | 330 | 5 270 | 27% 1.8 | 11% 2.004 | 61% 1.626 | 45% 1.703 | 1% 0.442 |
| 5 | Bakenov Zhumabay | 38 | 222 | 5 537 | 1% 0.35 | 58% 1.438 | 40% 2.08 | 25% 1.335 | 1% 0 |
| 6 | Issakhov Alibek | 33 | 261 | 4 063 | 24% 1.131 | 29% 2.704 | 41% 3.099 | 24% 1.836 | 5% 3.993 |
| 7 | Ramazanov Tlekkabul | 28 | 259 | 3 001 | 10% 0.646 | 34% 1.049 | 56% 0.597 | 2% 0.057 | 0 |
| 8 | Mun Grigoriy | 27 | 145 | 2 258 | 11% 0.367 | 30% 1.437 | 60% 0.852 | 4% 0.14 | 0 |
| 9 | Atabaev Timur | 25 | 94 | 1499 | 27 0.943 | 32% 1.498 | 24% 0.904 | 59% 1.131 | 10% 1.344 |
| 10 | Insepov Zinetula | 24 | 134 | 2 365 | 23% 0.887 | 52% 0.275 | 23% 0.457 | 48% 0.47 | 3% 0.391 |
| 90 | Abdullaev Azat | 11 | 43 | 306 | 32% | 0 | 68% | 12% | 0 |

| № | Author | H-index | DOC | CIT | FA/ FWCI1 | LA FWCI 2 | CoA/ FWCI 3 | CorA/ FWCI 4 | SA/ FWCI 5 |
|---|---|---|---|---|---|---|---|---|---|
| | | | | | 0.419 | | 0.212 | 0.515 | |
| *** | | | | | | | | | |
| 91 | Shunkeyev Kuanyshbek | 11 | 49 | 267 | 47% 0.667 | 28% 0.309 | 25% 0.653 | 9% 0.307 | 0 0 |
| 92 | Dzhumadildaev Askar | 11 | 66 | 462 | 79% 0.709 | 0 | 0 | 43% 0.514 | 21% 0.664 |
| 93 | Yesmakhanova Kuralay | 11 | 41 | 338 | 28% 2.487 | 16% 3.756 | 56% 1.512 | 6% 4.624 | 0 |
| 94 | Kenzhin Yergazy | 11 | 37 | 341 | 0 | 35% 0.966 | 65% 0.915 | 0 | 0 |
| 95 | Rakhadilov B. K. | 11 | 106 | 384 | 50% 0.828 | 5% 0.517 | 42% 1.327 | 12% 0.416 | 0 |
| 96 | Tuleushev Yu G. | 11 | 77 | 416 | 26% 0.123 | 19% 0.507 | 33% 0.267 | 44% 0.14 | 0 |
| 97 | Dosbolayev Merlan | 11 | 62 | 444 | 20% 0.394 | 3% 1.312 | 78% 0.662 | 15% 0.409 | 0 |
| 98 | Shunkeyev Kuanyshbek | 11 | 49 | 267 | 47% 0.667 | 28% 0.309 | 25% 0.653 | 9% 0.307 | 0 |
| 99 | Abdullaev Azat | 11 | 43 | 306 | 32% 0.419 | 0 | 68% 0.212 | 12% 0.515 | 0 |
| 100 | Tulepov Marat | 11 | 44 | 298 | 23% 0.446 | 14% 0.33 | 53% 1.357 | 12% 2.009 | 0 |
| 101 | Touzelbaev Maxat | 11 | 29 | 886 | 100% 1.283 | 0 | 0 | 0 | 0 |

where

DOC - total number of publications indexed in the SCOPUS database,

CIT - total number of citations for publications and the author,

FA – first author in the publication,

FWCI – average Field-Weighted-Citation-Impact (FWCI), a citation impact indicator weighted by subject area. The numbering corresponds to the author's role in the publication,

LA - last author in the publication,

CoA - coauthor (in places other than FA and LA),

CorA - corresponding author,

SA - single-authored (without coauthors).

Kazakhstani authors in the field of natural sciences most often appear as "coauthors." The second is the "last author," and the third is the "first author." It is noteworthy that there are almost no "single-authored" publications among Kazakhstani scientists on this list.

Note that in some fields or publications, authors may be listed in alphabetical order. However, according to the ranking of authors based on SCOPUS data (Article), there is no alphabetical order in the Kazakhstani collaboration. Nevertheless, in some (individual) cases, there is an alphabetical order of authors in international collaborations within scientific publications.

Let us calculate the correlation index between the parameters of Table 1:

*Pearson correlation coefficient between H and FA - 0.326,*

*Pearson correlation coefficient between H and LA - 0.271,*

*Pearson correlation coefficient between H and CoA - 0.177,*

*Pearson correlation coefficient between H and CorA - 0.216.*

Note the average negative correlation (-0.326) between the Hirsch index of an individual scientist and their participation in collaborative publications as the first author (FA) (other correlations, being weakly negative, are not considered).

Based on the graph showing the dependency trends of the author's functional participation in publications on the Hirsch index, we will create trend graphs:

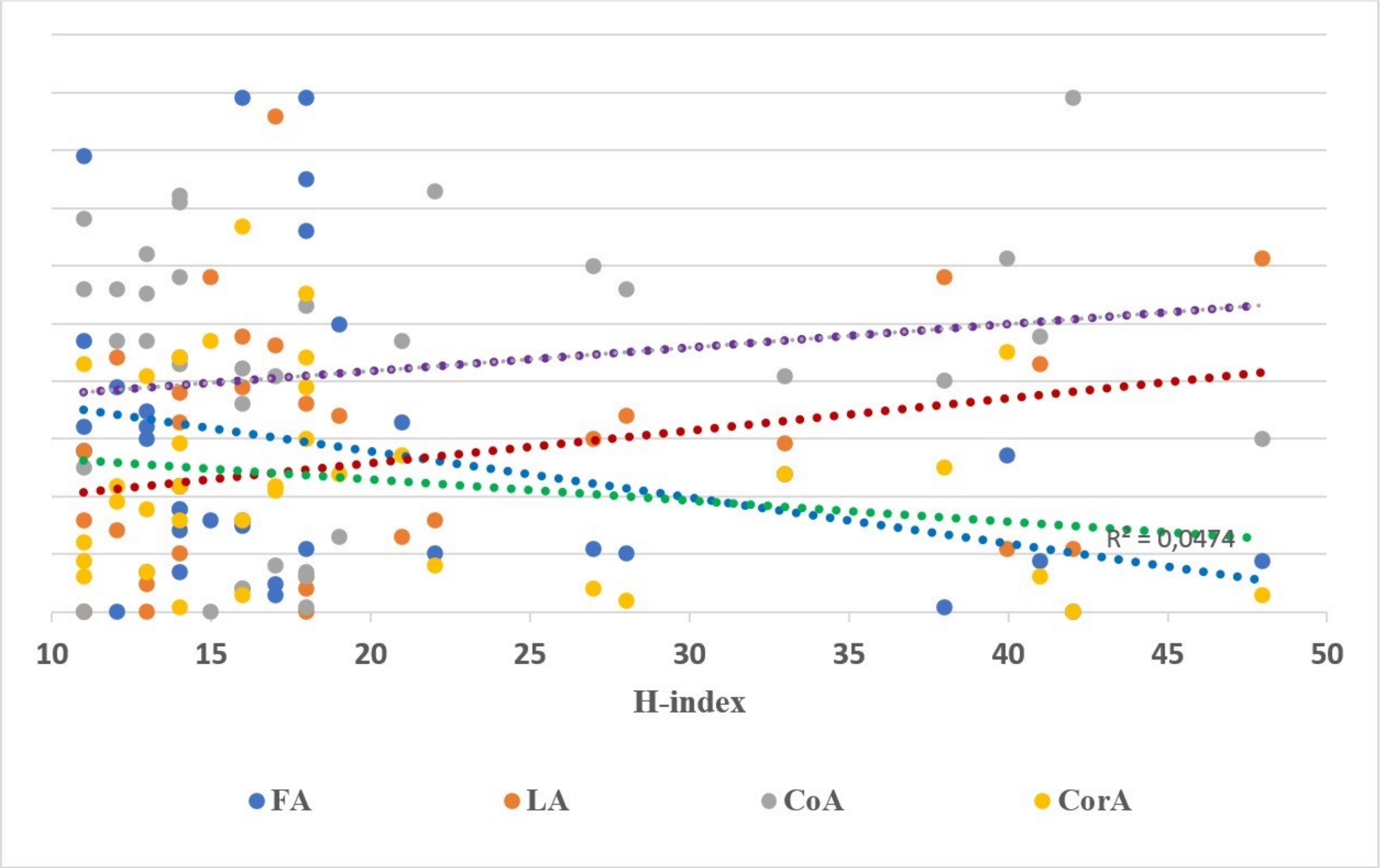

**Fig 1.** Trends in the dependence of FA, LA, CoA, and CorA on the H-index

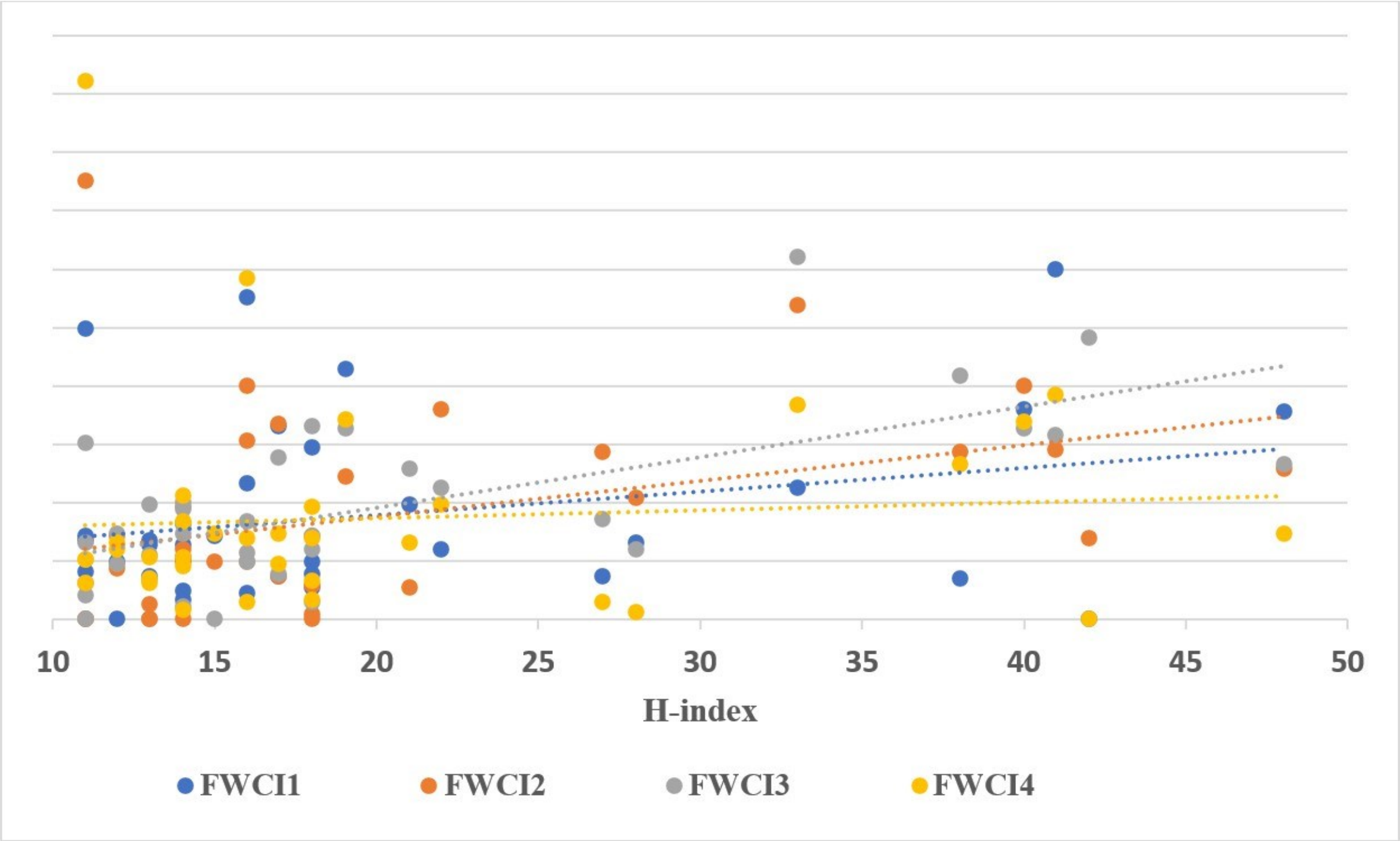

**Fig. 2.** Trends in the Dependence of the Average FWCI on the H-Index

The trend lines in graphs 1 and 2 correspond to the degree of active and passive participation of the author in collaborative publications.

Two distinctly identical trends are highlighted:

*Trend 1: As the Hirsch index increases, the degree of participation of Kazakhstani authors in collaborative (international) articles, interpreted through the "First Author" and "Author-Correspondent," decreases.*

*Trend 2: As the Hirsch index increases, the degree of participation of Kazakhstani authors in collaborative (international) articles becomes more diffuse (through the diffusion of active participation), as interpreted through the "Last Author" and "Coauthor."*

We consider and analyze these trends only as a hypothesis (based on low approximation values). To confirm this hypothesis, it is also necessary to examine the simultaneous dynamics of not only the entire group of researchers but also the chronological dynamics of each individual scientist from an H-index of 10 to their current maximum within the framework of the ergodic hypothesis.

If we look at individual temporal profiles with a high Hirsch index, in certain cases, an identical regularity is observed:

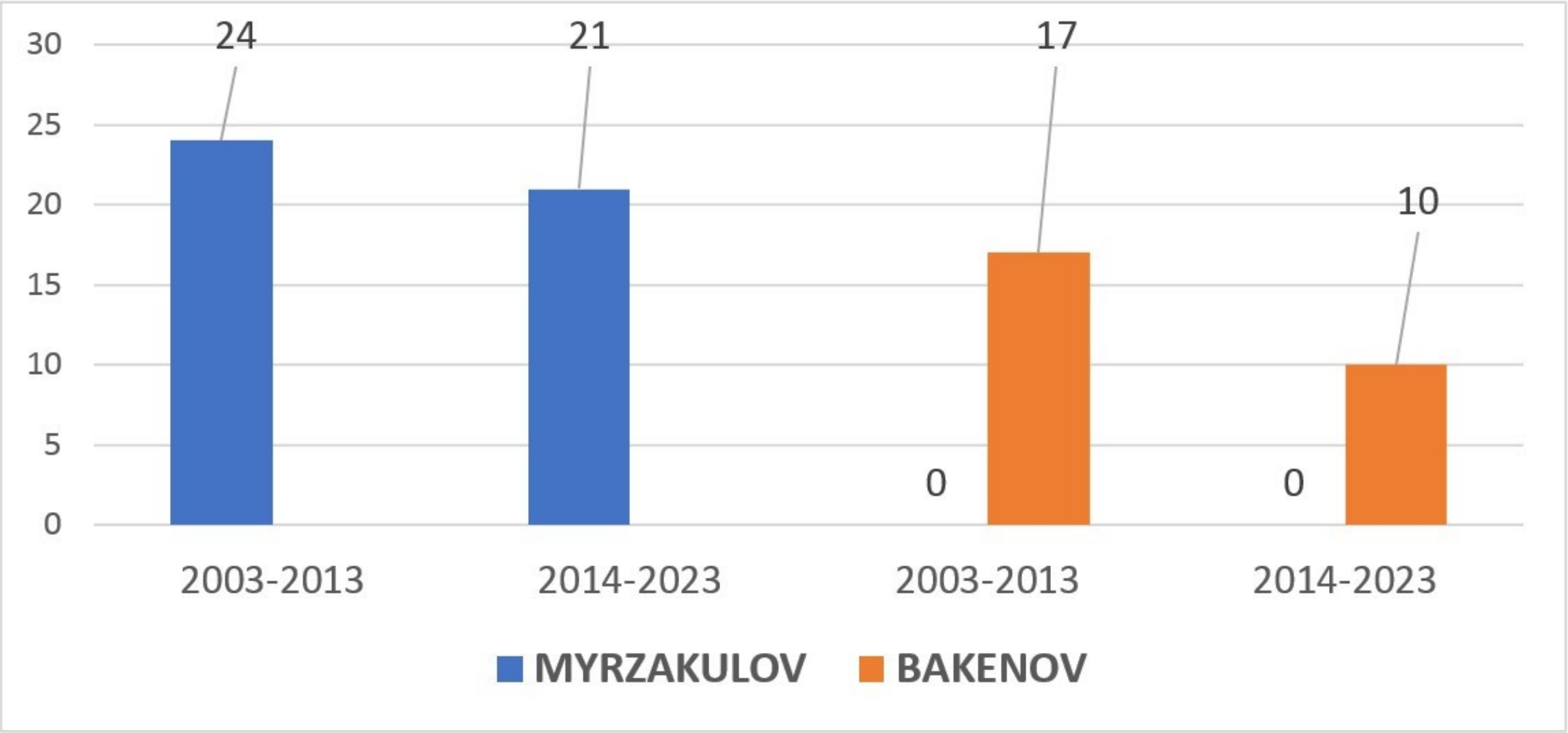

**Fig 3.** Dynamics of the "First Author" position for individual researchers

The "first author" parameter for Ratbay Myrzakulov from 2003 to 2013 was 24 (data from Google Scholar). However, for the period 2014-2023 inclusive, it is 22. The number of "first author" parameters for Zhumaabay Bakenov from 2003 to 2013 inclusive is 17. However, for the period 2014-2023 inclusive, it is 10.

This allows us to formulate a hypothesis about the ergodicity of the scientific community in Kazakhstan in the context of publication activity and the role functions of coauthorship.

As the participation of Kazakhstani authors in joint international publications increases, their level of activity decreases due to the replacement of corresponding activities by coauthors. Predominantly, these coauthors are foreign scientists with a higher Hirsch index.

Effectively, the foreign coauthor "delegated" a part of their Hirsch index to the Kazakhstani scientist (within the framework of synergy), taking on the functions of the first author. In other words, taking on a more active role in shaping the substantive scientific part of the joint publication.

This aligns directly with research showing a positive (patronage) influence of international scientific collaboration through joint publications on the quality and effectiveness of national sciences.

This indirectly confirms the weak positive correlation (0.177) between the Hirsch index and the increase in Kazakhstani coauthorship (CoA) in joint international publications.

Additionally, according to SCOPUS statistical data, in joint publications, the level of international participation usually increases against the background of an increase in the Hirsch index of Kazakhstani researchers.

Thus, as a hypothesis, a substantial increase in the H-index of Kazakhstani scientists is noted as the degree of scientific localization (national participation) decreases and the degree of "foreign content" increases in knowledge-intensive content within the framework of international collaboration.

## 3. K-index

As noted above, the percentage of "single authorships" in the scientific publications of leading Kazakhstani scientists is negligible. Accordingly, the question of coauthorship (predominantly international coauthorship) in science requires special attention. The number of joint publications (especially at the international level) and the percentage of joint international publications in the total number of publications are growing dynamically

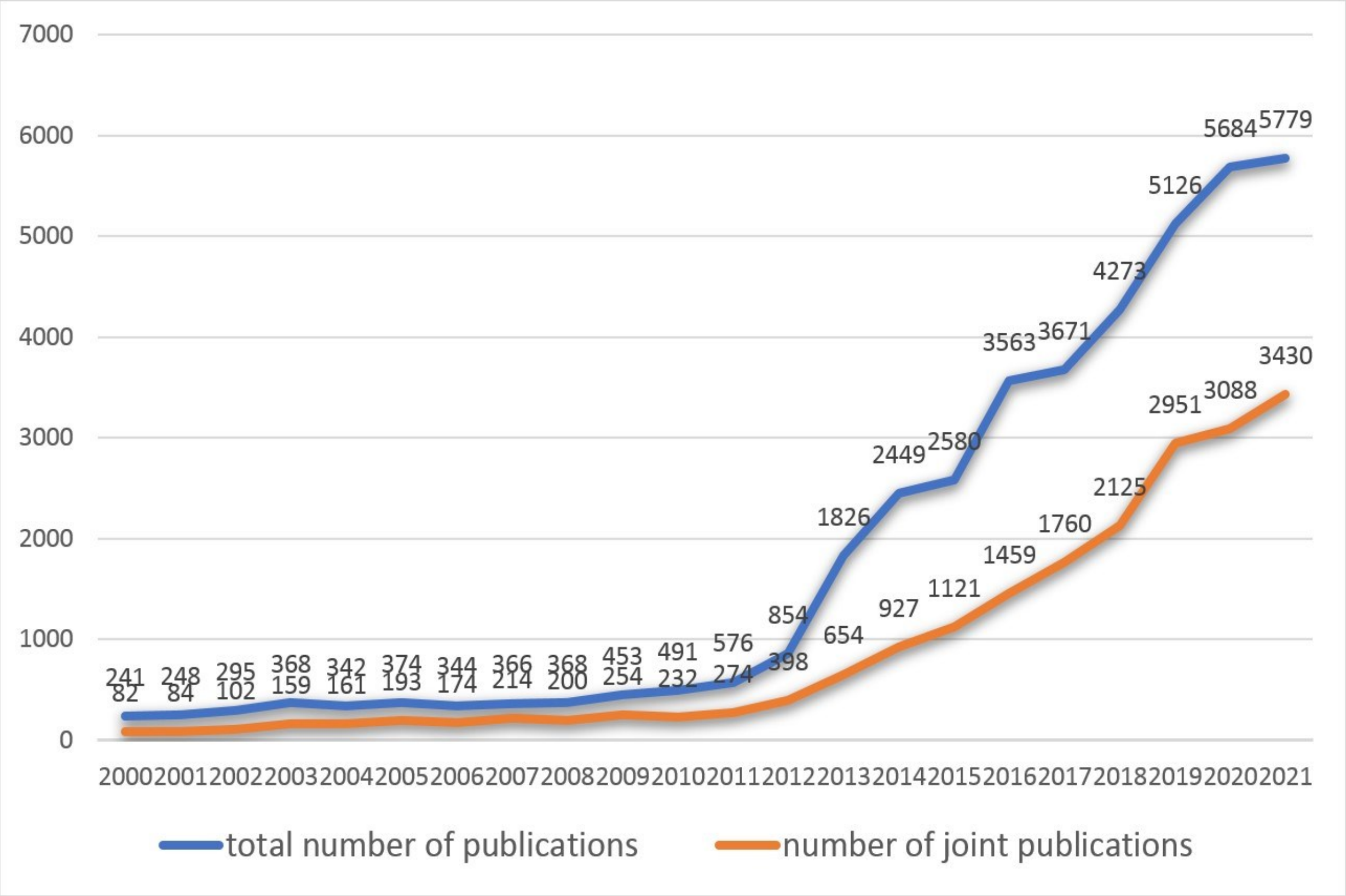

**Fig4.** Comparative dynamics of the total number of publications and the number of joint publications (data from *SCImago Journal & Country Rank* for Kazakhstan)

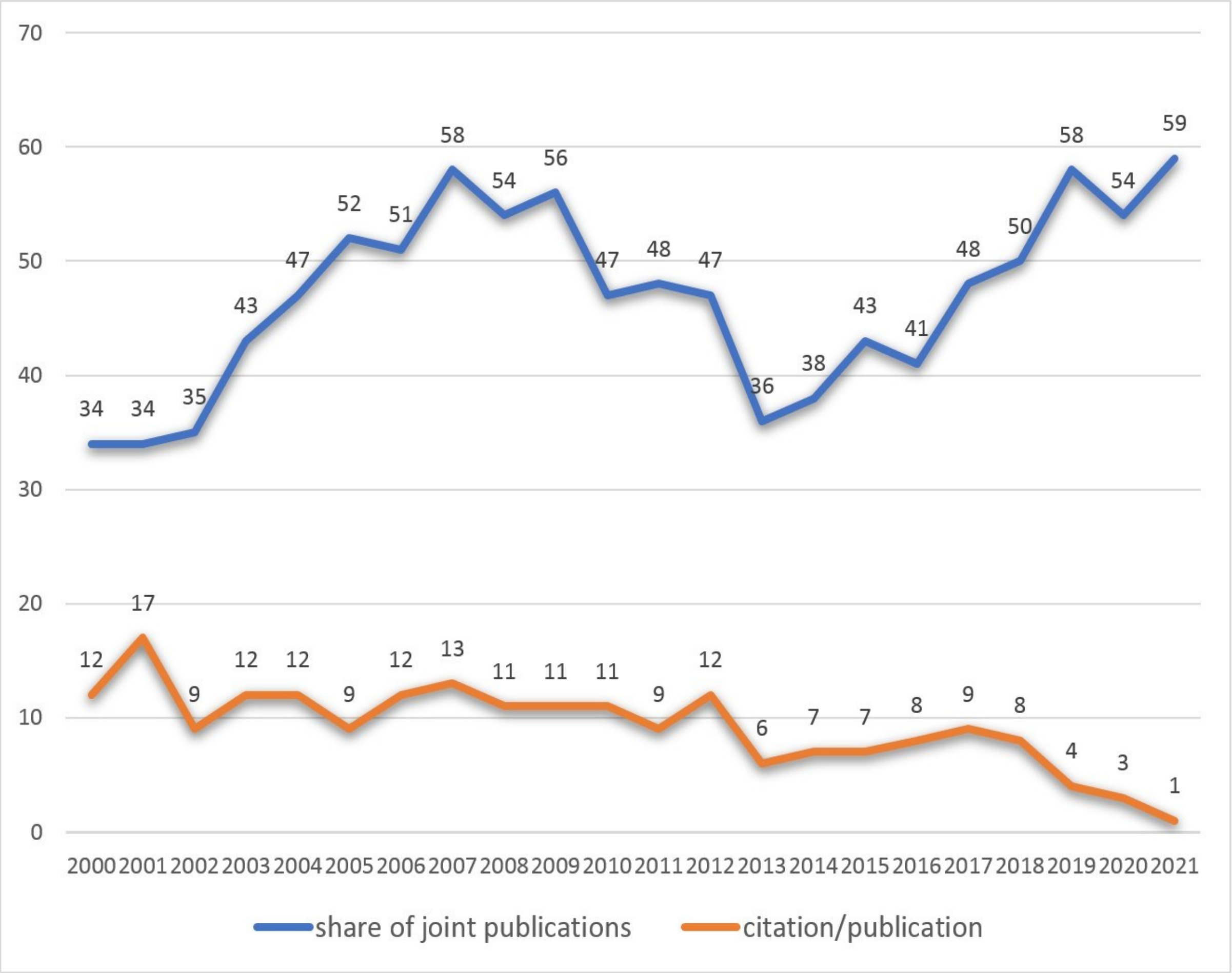

**Fig 5.** Ratio of the share of joint publications (%) to the total volume and the number of citations per publication (for Kazakhstan)

Typically, the number of coauthors in joint publications ranges from 2 to 4 scientists [Adams, 2021, p. 325], but it can exceed 100 coauthors (hyperauthorship).

Therefore, the assessment of the quality and effectiveness of scientific activity can be approximated through the collaborative publication activity of coauthors.

A joint scientific publication can be interpreted as a cooperative (coalitional) game with zero sum (von Neumann, 1944), in which the payoff function is determined through the hierarchy of roles of coauthors in the joint publication.

We consider the joint publication activities of four coauthors, A, B, C, and D, from the perspective of their role functions:

FA - the first author,

CorA - corresponding author,

CoA - coauthor,

LA - the last author,

SA - single author.

The role functions FA and CorA are winning functions since they provide an advantage compared to the remaining functions in terms of the overall citation and mentionability of the coauthor. In general, the role functions FA and CorA reflect the maximum comparative contribution of the coauthor to the scientific value of the publication. The CoA and LA functions are losing function because they minimize the roles of the author in the publication and their mentionability (in the context of total citations). The role functions FA and CorA make comparatively smaller contributions to the scientific value of the publication (but not always).

The overall payoff function of a joint publication can be written as the sum of the individual payoff functions of the four coauthors:

F(i) = $\sum_{i=1}^{4} f(i)$ = $f(A)+f(B)+f(C)+f(D)$

From the set of alternative mixed role strategies for the four coauthors,

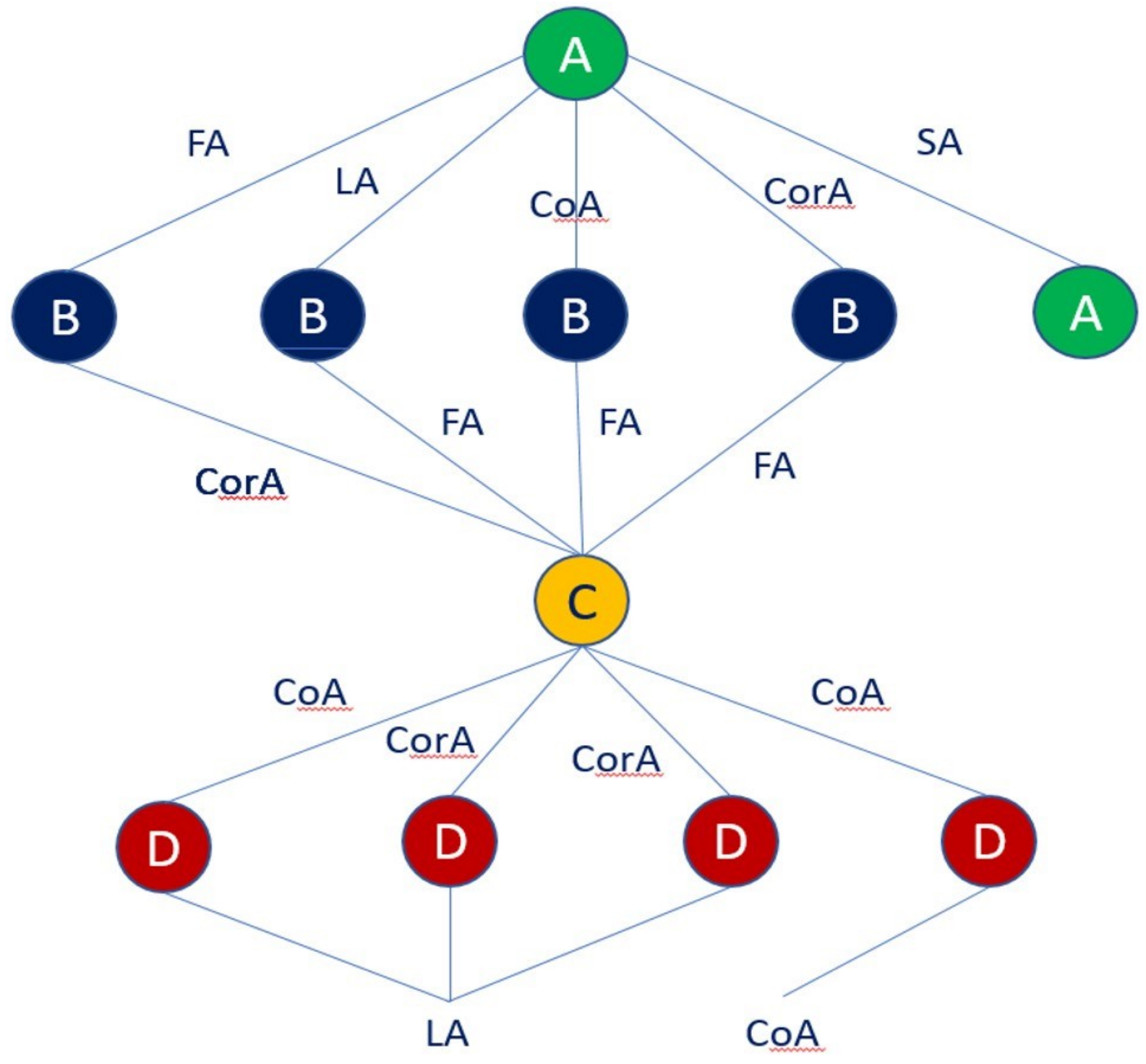

**Fig.** 6. Alternative (mixed) strategies of role functions in coauthorship

Let us consider the dominant winning (pure) strategies of coauthors:

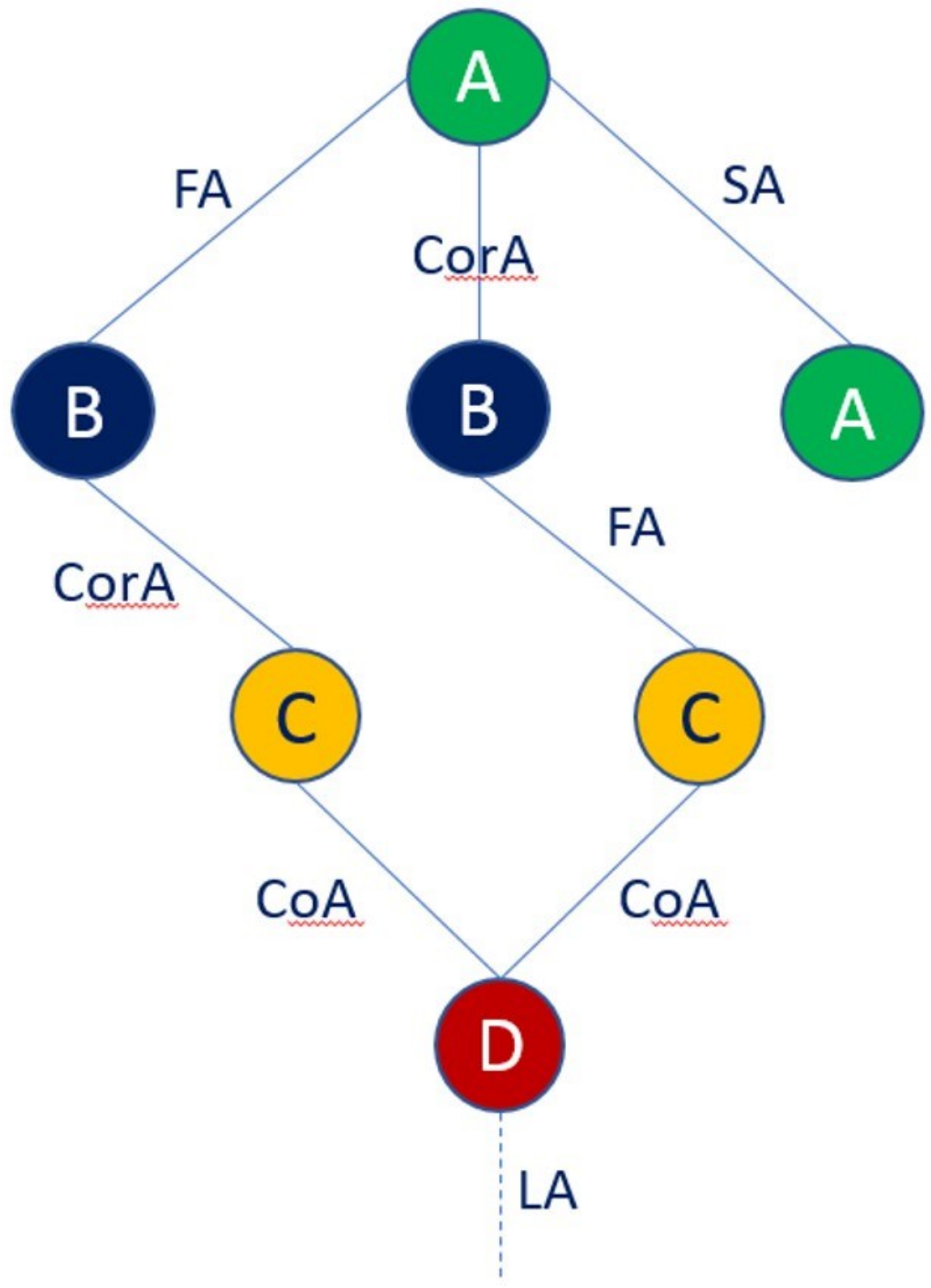

**Fig**. 7. Winning (pure) strategies of role functions in coauthorshi

In the cooperative game of collaboration (without SA), it is not mandatory that coauthor A will choose the winning strategy FA with a probability of 100%. The CorA strategy has positive effects on communication and the establishment of positive feedback.

Strategies involving CoA and LA are considered conditional loss strategies. Coauthor C will choose the CoA strategy with a higher probability, but not equal to zero (for example, in the case of alphabetical order of coauthors). Coauthor D will be forced to choose the LA strategy with the maximum probability as the probabilistically only remaining option.

The winning strategies of the dominant role functions can be presented as:

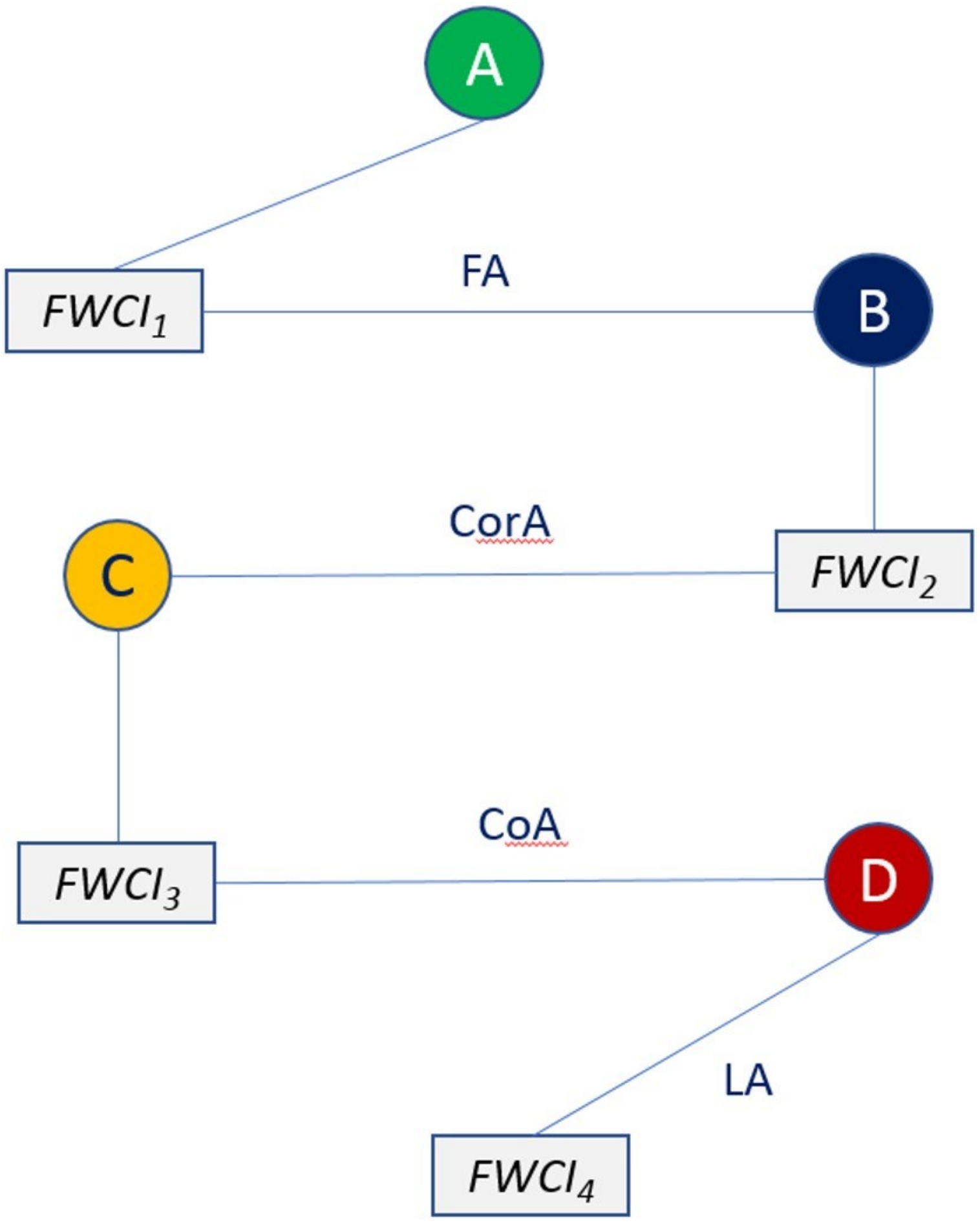

**Fig**. 8. Dominant strategies of role functions in coauthorship

Here, the parameter FWCIi (i=1,2,3,4) indicates how winning a particular role function was in the final set of corresponding publications in terms of citation impact. Let us introduce the role dominance coefficient kr, reflecting an individual contributor's impact on the scientific value of a publication, in the following way:

$$k_r = \frac{1+(FA + CorA + SA)}{1+(CoA+ LA)} \tag{1}$$

The winning function in the role strategy is represented as:

$$\sum_{i=1}^{4} f(i) = k_r * \mathrm{FWCI}$$

where

$$\mathrm{FWCI} = \sum_{i=1}^{4} FWCI(i) = \mathrm{FWCI}_1 + \mathrm{FWCI2} + \mathrm{FWCI}_3 + \mathrm{FWCI}_4$$

In this case, the FWCI is the sum of the average FWCI values (field-weighted citation impact).

Now, let us consider the second winning function—citation per document—as a noncooperative game with a nonzero sum. The process of citing a publication in the context of the publication-perception interaction can be interpreted as a noncooperative game with a nonzero sum.

The citation tree in the "publication-perception" system can be represented as follows:

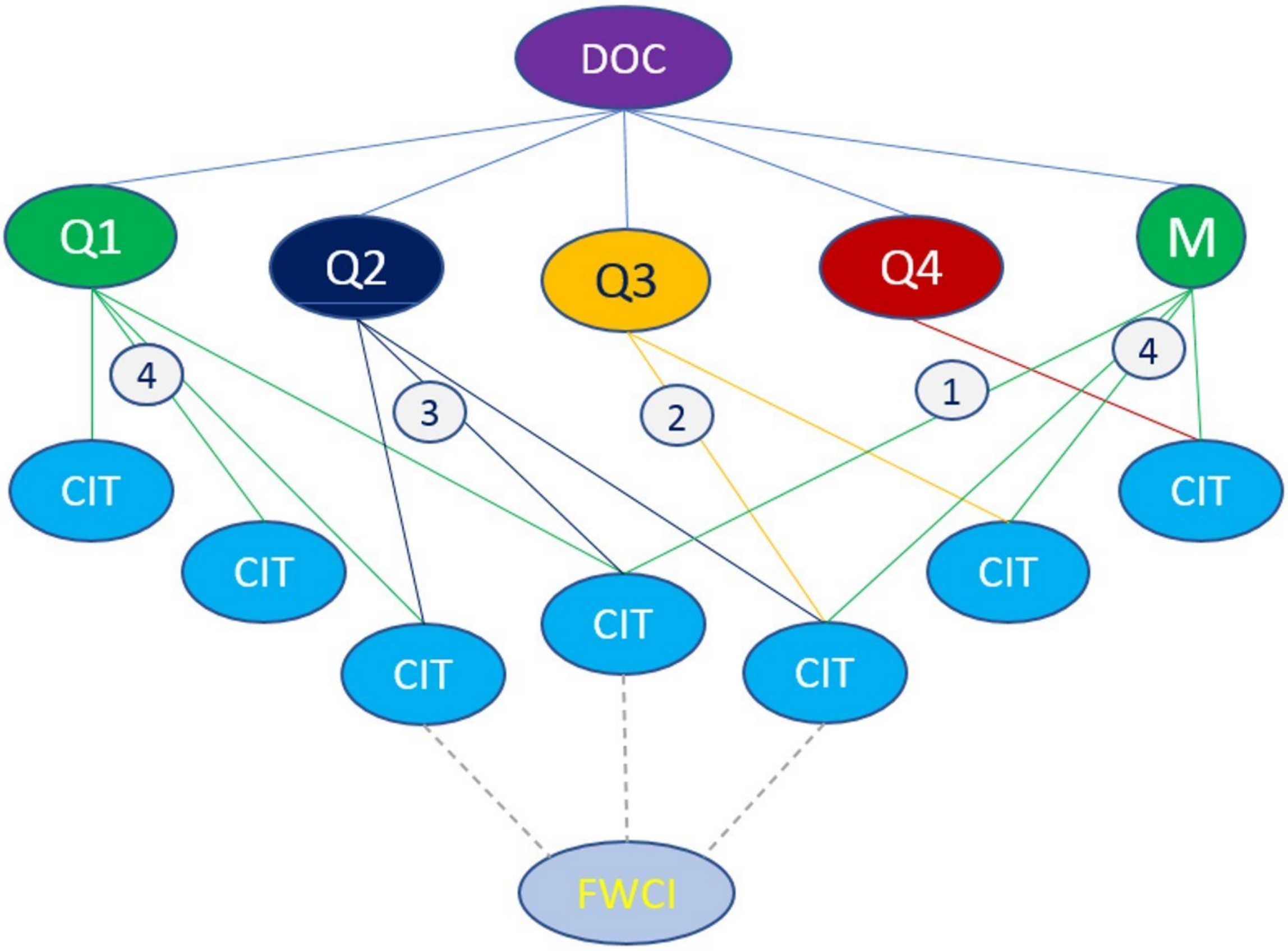

**Fig**. 9. Citation tree CITs of publication DOCs according to journal quartile

A document can be published in a journal with one of the quartiles Q1, Q2, Q3, or Q4 or as a print-electronic book publication (monograph, scientific manual, book, treatise). The probability of citing a publication depends on its scientific value (demand) and the quartile of the scientific publication. In Q1 journals, the citation probability is 4 conditional units; in Q2, it is 3 conditional units; and so on.

The number of potential users/perceivers is greater for journals with a higher quartile. Additionally, the Matthew effect works here: the same publication is more likely to be cited in a Q1 journal than in a Q4 journal.

The value of a publication in this sense is interpreted indirectly through the quartile and directly through the citations per publication.

The correction for scientific fields and directions is interpreted through FWCI – the average Field-Weighted-Citation-Impact (a citation impact indicator weighted by subject area).

In matrix form, winning strategies based on scientific value are interpreted through citation parameters per unit of scientific output:

$$F(cit) = \begin{bmatrix} CIT & 0 \\ 0 & 1/DOC \end{bmatrix} = \frac{CIT}{DOC}$$

Here, $F(cit)$ is a second-order diagonal matrix representing the pure citation strategy of a coauthored publication, and $CIT/DOC$ is the parameter expressing the value (demand) of scientific output in relative terms.

The hybrid evaluation strategy of the quality and performance of a scientist $F(s)$, in terms of publication activity interpretation, will be a superposition (linear function) of role function parameters and citation impact:

$$F(k)=F(i)+F(cit)$$

These parameters are considered fundamental indicators of the scientific significance of a researcher and the value of a scientific publication. In this case, the heuristic function for evaluating the quality and performance of an individual researcher is represented as:

$$K = k_r FWCI + \frac{CIT}{DOC} \qquad (2)$$

Here, *K* is the aggregated index of the quality and performance of scientific activity, interpreted through a researcher's contribution to a publication and citation as scientific value. In this case, the multiplier (*kr·FWCI*) represents a second-order small correction coefficient.

We call this index the K-index, the Kazakhstani index of scientific citation. It reflects the quality and performance of publication activity, specifically in research, without tying it to activities in commercialization or other applied sciences where quality and performance are interpreted differently and may be considered.

In this context, considering patent activity, commercialization, and economic impact, the integrated Kazakhstani index of scientific citation is presented as follows:

$$K_i = \sum_{i=1}^{3} K(i) = K + K_p + K_c$$

where:

*Ki* is the integrated K-index,

*K* is the K-index,

*Kp* is the patent activity index according to the regulations of Kazpatent (National Institute of Intellectual Property) and WIPO (World Intellectual Property Organization),

*Kc* is the commercialization index (economic impact based on ROI (Return on Investment) criteria.

## 4. K-index ranking

Now, for example and analogous to Table 1, let us present the K-index table for individual researchers:

**Table 2.** K-rating: K-index values for 50 Kazakhstani researchers (based on SCOPUS data only).

| № | Researcher | CIT/DOC | WFCI | $k_r$ | K-index |
|---|---|---|---|---|---|
| 1 | Konarov Aishuak | 51,83 | 8 | 0,75 | 58 |
| 2 | Zhautykov Bulat | 45,33 | 3,1 | 0,5 | 47 |
| 3 | Abdikamalov Ernazar | 35,55 | 4,6 | 0,69 | 39 |
| 4 | Utepbergenov D. | 35,5 | 2,1 | 0,94 | 37 |
| 5 | Boranbayev Askar | 11,68 | 8,2 | 2,37 | 31 |
| 6 | Myrzakulov Ratbay | 26,41 | 5,1 | 0,57 | 29 |
| 7 | Bakenov Zhumabay | 24,94 | 5,2 | 0,64 | 28 |
| 8 | Shaĭkenov Block | 27 | - | - | 27 |
| 9 | Issakhov Alibek | 15,57 | 12,76 | 0,9 | 27 |
| 10 | Kenessov Bulat | 23 | 3,57 | 0,79 | 26 |
| 11 | Konuspayeva Gaukhar | 20,48 | 4,18 | 0,89 | 24 |
| 12 | Kozlovskiy Artem | 15,97 | 7,6 | 1 | 24 |
| 13 | Kudaibergenov Sarkyt | 17,23 | 6,05 | 0,96 | 23 |
| 14 | Atabaev Timur | 15,95 | 5,8 | 1,26 | 23 |
| 15 | Omarov Rustem | 20,3 | 2,77 | 0,68 | 22 |
| 16 | Turuspekov Yerlan | 19,58 | 2,58 | 0,99 | 22 |
| 17 | Sarsenbekuly Bauyrzhan | 18,51 | 5,7 | 0,57 | 22 |
| 18 | Sypabekova Marzhan | 18 | 5,89 | 0,72 | 22 |
| 19 | Sarsenbi Abdizhakhan | 14,8 | 6,26 | 1,13 | 22 |
| 20 | Aidarova Saule | 18,67 | 4,5 | 0,62 | 21 |
| 21 | Zdorovets Maxim | 15,94 | 7,97 | 0,59 | 21 |
| 25 | Zayadan Bolatkhan. | 18,32 | 3,12 | 0,66 | 20 |
| 26 | Insepov Zinetula | 17,65 | 2,48 | 0,99 | 20 |
| 27 | Mentbayeva Almagul | 14,67 | 6,8 | 0,77 | 20 |
| 28 | Jumabekov Askhat | 14,55 | 6,2 | 0,9 | 20 |
| 29 | Askarova Aliya | 10,6 | 5,65 | 1,73 | 20 |
| 30 | Moldabekov Zhandos | 13,65 | 6,26 | 0,82 | 19 |
| 31 | Sadybekov Makhmud | 9,8 | 7,64 | 1,2 | 19 |
| 32 | Nurkeeva,Zauresh | 17,83 | (1) | 0.5 | 18 |
| 33 | Dzhumagulova Karlygash | 15,17 | 3,2 | - | 18 |
| 34 | Mun Grigoriy | 15,57 | 2,8 | 0,6 | 17 |
| 35 | Matkarimov Bakhyt | 14,6 | 2,04 | 0,596 | 16 |
| 36 | Roman'kov S | 12,85 | 1,77 | 2 | 16 |
| 37 | Torebek Berikbol | 8,79 | 8,68 | 0,8 | 16 |
| 38 | Ogay Vyacheslav | 13,7 | 1,8 | 0,8 | 15 |
| 39 | Ustimenko Alexandr | 12,48 | 2,65 | 1 | 15 |
| 40 | Umirbaev Ualbay | 12,1 | 2,44 | 1,16 | 15 |
| 41 | Kalmenov Tynysbek | 11,23 | 3,13 | 1,3 | 15 |
| 42 | Ramazanov Tlekkabul | 11,59 | 2,3 | 0,59 | 13 |
| 43 | Kenzhina Inesh | 11,36 | 3,1 | 0,6 | 13 |
| 44 | Suleimen Yerlan | 7,96 | 2,85 | 1,72 | 13 |
| 45 | Folomeev Vladimir | 10,12 | 2,58 | 0,68 | 12 |
| 46 | Dzhumadildaev Askar | 7 | 1,89 | 2,43 | 12 |
| 47 | Sassykova Larissa | 6,48 | 4,9 | 1,215 | 12 |
| 48 | Bolegenova Saltanat | 10,25 | 2.23 | 0,51 | 11 |
| 49 | Kadyrzhanov Kayrat | 8,3 | 3,96 | 0,64 | 11 |
| 50 | Dzhunushaliev Vladimir | 7,64 | 1,84 | 2 | 11 |

Thus, the K-index reflects not publication activity (according to Hirsch) but the publication efficiency of a researcher.

Let us consider the position of the hypothetical leader in the ranking—Aishuak Konarov (Assistant Professor, School of Engineering and Digital Sciences, Department of Chemical and Materials Engineering, Nazarbayev University). He has a total of 41 publications. His recent articles in journals such as Small (Q1), Electrochemistry Communications (Q1), Advanced Energy Materials (Q1), and RSC Advances (Q2, was Q1 until 2022) have attracted significant attention.

His total number of citations was 2,153, averaging more than 50 citations per publication. This figure is the highest among Kazakhstani scientists. In 28% of Konarov's publications, he served as the "first author" (FA). For publications where Konarov is the "first author" (FA), the average number of citations exceeds the industry average by five times (average FWCI = 5.196). This indicates that the citation efficiency of his publications in his scientific field is five times greater than that of others, on average, and so on for other role functions.

Thus, Konarov's publication efficiency is the highest among Kazakhstani scientists within the calculated list.

Bulat Zhautykov (Satbayev University) takes second place in the K-rating, also holding second place in the Hirsch index list.

Several trends can be observed from the table:

1. The K-index places emerging, young, and relatively young scientists at the top who currently have a small number of publications, but these publications have a high average citation coefficient, indicating demand and involvement. Moreover, the average citation rate in the industry and dominant coauthorship role functions also have high values.
2. At the same time, scientists with a high Hirsch index continue to maintain stable leading positions in the K-index rankings, although not necessarily at the very top.

In conclusion, the K-index introduces the principles of inclusivity in scientific activity evaluation, moving away from the extractive principle of the Hirsch index. Additionally, it introduces the principles of recursion, where the K-function calls itself to identify a feature, making it a recursive and iterative procedure simultaneously.

The K-index does not reflect scientific insignia (awards, honors, diplomas, etc.), but for a comprehensive picture, it can be modified to incorporate them.

Another noteworthy point is that there are no anomalous outliers in the K-index ranking table. All scientists who held leading or top positions in Table 1 (according to the Hirsch index) also occupy top or leading positions in Table 2 (according to the K-index). Therefore, the K-index is relevant to the Hirsch index (and vice versa). The specific positions of researchers are simply refined (adjusted) in terms of the K-index.

Positive aspects of the K-index can be expressed as follows:

a) The index does not require the continuous (avalanche) production of publications.
b) The index prioritizes the quality of an individual publication based on its citation coefficient (demand).
c) The index takes into account each scientific publication (unlike the H-index).
d) The index considers the degree to which a researcher participates in a scientific publication.
e) The index (similar to the Hirsch index) only considers indexed publications.

The K-index is deliberately not tied to percentiles and journal quartiles but is linked to their indexation through SCOPUS and other databases. This inertia allows Kazakhstani scientists not to overly adhere to the quartile of journals, focusing on the quality of the material rather than the journal's prestige. Of course, this does not exclude but implies a preference for publications in high quartile (Q1-Q2) and high percentile journals in each quartile, according to "Figure 2. Dominant Winning (Pure) Strategies of Coauthorship Role Functions."

Additionally, the condition of indexing publications in SCOPUS or WoS for the K-index automatically eliminates the need for publications in "predatory journals." In the most extended version, the index may consider the quartile and percentile of journals in which the author publishes through the Qn coefficient, which takes into account the average quartile and percentile for all or selected recent publications.

In cases where an author ranks in a particular scientific field or narrow specialization occurs in alphabetical order, the role dominance coefficient is taken as 1.

The use of the K-index will address the main problem in modern science—the overwhelming and often unjustified flood of publications. This is achieved by prioritizing quality and scientific activity over quantity and mechanical coauthorship

**Table 3**. Parameters and characteristics of the K-index.

| **K-INDEX** | |
|---|---|
| PROPERTY | NOTE |
| In the index, the principle of publication activity is shifted to the principle of publication efficiency | Nonquantitative but qualitative indicators of the final scientific product |
| Scientific recognition according to the index can come much earlier than with the Hirsch index. Since it does not require an accumulation period of materials | The K-index is a recursive function, whereas the H-index is an iterative function |
| The index equally interprets both journal and book publications (monographs, textbooks, treatises, books) | Ideally, having one book-monograph can secure leading positions in the ranking by the index |
| The index is indifferent to the number of publications, considering quantity as a relative parameter | The quantity of publications should not be considered an absolute parameter of the quality and effectiveness of scientific research |
| The index depends on the talent of individual scientific work and the degree of the researcher's participation in that work | The index is optimal for young scientists who, for talented or even ingenious initial 1-3 works, can receive well-deserved recognition without transitioning to the stage of the so-called "productivity according to Hirsch" of articles to increase the Hirsch index |
| The index is invariant concerning the chronology of a scientist's scientific activity | Recognition through the index can come at the age of 18 or 81. The index takes into account but does not recognize seniority |
| The index considers the degree of active and passive role participation of a researcher in the publication | The degree and form of participation of the scientist in the publication (project) are important parameters of his scientific activity, which should be taken into account |
| The index is maximally universal by considering the specific characteristics of individual scientific disciplines | The consideration is interpreted through FWCI, which has its specific relative value for each scientific direction or field |
| The index can change not only toward an increase but also toward a decrease | This allows evaluating or determining the scientific potential of a scientist at the current or demanded moment. Moreover, it allows examining the chronological dynamics of his scientific activity and performance |
| The index is sensitive to the number of coauthors | In the interpretation of role functions, which depend on the number of coauthors |
| All data embedded in the index are easily verifiable, reproducible, identifiable, and correct within the framework of open statistics presented in the SCOPUS databas | To assess the K-index, there is no need for the presence of paid license or analytical tools for processing indexed databases. |

Here, it is necessary to clarify several points related to the sufficient and necessary criteria of the K-index.

First, the following question arises: how can we account for an author's contribution to a publication with multiple coauthors (poly- and hyperauthor publications where the number of coauthors exceeds 4)?

a) In polyauthor publications, the Ringelmann effect (social loafing) (Ingham, 1974) can manifest itself. In collective work, a person may invest less effort than in individual work. Therefore, the Ringelmann coefficient can be introduced into the index in the following original (or interpreted) form: $S = 100 - 7 (K-1)$, where S is the average individual contribution of a coauthor as a percentage and K is the number of coauthors. However, it is essential to consider that scientific work is primarily intellectual rather than physical labor. The intellectual contributions of individual scientists in joint research or publications are not additive; rather, they are synergistic (ideally). Even in a publication with 100 coauthors, the scientifically intensive (intellectual) contribution of an individual author can reach 100%.

b) Second, the adjacent Zajonc effect (Zajonc, 1993) often manifests in joint scientific publications. In collaborative publications, the presence of coauthors often stimulates increased intellectual and physical dedication from each coauthor. This can be an additional factor of synergy in collective research or joint publication. In polyauthor collectives, a scientist may work more or better than in individual work.

c) Third, according to general systems theory (von Bertalanffy, 1962), "the whole is more than the sum of its parts." The same principle applies to the synergy of work. The result of the work significantly differs from the simple sum of the contributions of individual scientists. However, how much synergy is achieved depends on the citation of the publication or the popularity of the scientific research.

Therefore, it is not necessary to account for the number of coauthors in the index through a separate corrective coefficient. The contribution of a coauthor to a polyauthor publication is normalized through the coauthorship role parameter (CoA).

The following question arises: is there a need to consider insignia (titles, degrees, awards, etc.) in the index?

Here are some considerations: degrees and titles are the result of scientific activity (not always, but often). However, scientific results and achievements are not the result of degrees and titles (always and unconditionally), especially in the Kazakhstani scientific system. The assessment of scientific potential should be inclusive, not extractive. Awards and other honors are already interpreted through citations (including the Matvey effect).

In conclusion, indices should only consider scientific results, not scientific or quasiscientific achievements. Evaluate solely based on scientific contribution and results, not on titles, degrees, regalia, or scientific longevity – the main principle of the K-index.

Therefore, introducing additional coefficients for the K-index is not needed. However, the "regalia index" can be considered in personnel matters or in "honorary titles."

Of course, if one absolutes purely quantitative indicators, no index can be universally applicable to all scientific fields. According to Goodhart's Law (Goodhart, 1975): "When a measure becomes a target, it ceases to be a good measure." Although the K-index is maximally universal in this regard, a grading table of the K-index (or another index) can be compiled separately for each scientific field or direction.

## 5. Conclusion

In contemporary epistemology, "collaboration in research positively influences publication productivity" (Lee, 2005), which is especially relevant in the current era of international science (Adams, 2013).

It is essential to emphasize that collaboration does not always lead to an increase in the scientific significance of research, considering the transactional, communication, and confrontational costs associated with research conducted by parties with different potentials and capabilities. However, overall, forms of collaboration often have a positive impact on research productivity (Abramo, 2017).

Although collaboration itself does not always have a positive effect on international collaboration, we can state that it exhibits signs and properties of commensalism using biological terminology.

The encyclopaedia Britannica defines commensalism as follows:

"*Commensalism - in biology, the relationship between individuals of two species in which one species obtains food or other benefits from the other without harming or benefiting the latter.*"

International scientific collaboration (IRC – International Research Collaboration) is undoubtedly also beneficial for developing or catching up to national scientific systems (Scarazzati, 2019), as such systems prefer collaboration with more advanced and developed systems.

However, commensalism in the interpretation and notation of the Hirsch index leads to significant costs in the functioning and development of some national scientific systems with a significant share of indigenous scientific directions and needs.

Jose Hirsch's publication strategy aims to maximize the author's utility function by increasing the number of publications, resulting in a corresponding increase in the number of citations. This strategy neglects other approaches aimed at improving the quality of individual publications or maintaining an active role in shaping scientific content.

Collaboration is a necessary condition for a high Hirsch index. Collaboration allows the quality function to be improved to increase the value of the quantity function. A single scientist can create one publication in a fixed period with 100 citations, resulting in a Hirsch index of 1. However, if they collaborate with as many coauthors as possible, they can generate 10 articles in the same fixed time interval, each cited 5-10 times. In this case, their utility function yields a Hirsch index of 5-10.

The strategy for the K-index is to maximize the author's utility function under Pareto optimality conditions, with the dominance of the quality of individual articles in terms of citations and minimizing the number of collaborators.

The total number of citations directly correlates with the overall number of scientific publications – this is a probability distribution. However, the "number of citations per publication" depends more on the quality – scientific value and significance – rather than the quantity of publications. If we abstract from the transition from quantity to quality, the number of citations per publication will depend more on the percentage of collaborative publications. That is, on the "scientific quality" of coauthors.

Since Kazakhstani scientists, a priori, are more oriented toward collaboration with researchers from more scientifically developed countries in joint publications, it can be assumed that the percentage of collaborative publications will directly affect the "number of citations per publication" (in the array of collaborative publications).

A researcher with a much lower H-index may have a higher K-index. The "publish or perish" principle remains relevant. However, this does not mean that one should produce as many publications as possible in the hope of surviving in the scientific field for as long as possible.

In addition to this life-affirming principle, another principle becomes relevant – the English "*Less is more*" or the German "*Besser wenig und gut als viel und schlecht*" (better little and good than much and bad).

As mentioned earlier, the necessity of a massive number of publications, driven by the dominance of the Hirsch index, leads to a decrease in the scientific content and quality of the publications:

**Table 4.** Overall scientific activity in Kazakhstan by year (SCImago Journal & Country Rank, data as of 08/21/2023)

| YEAR | DOC | CIT (DOC) | CIT | SelfCIT | CIT/DOC |
|---|---|---|---|---|---|
| 2000 | 242 | 240 | 3198 | 419 | 13.21 |
| 2001 | 248 | 244 | 4291 | 490 | 17.30 |
| 2002 | 295 | 290 | 2905 | 593 | 9.85 |
| 2003 | 368 | 361 | 4595 | 0 | 12.49 |
| 2004 | 342 | 338 | 4248 | 629 | 12.42 |
| 2005 | 373 | 371 | 3727 | 629 | 9.99 |
| 2006 | 344 | 336 | 4385 | 639 | 12.75 |
| 2007 | 369 | 363 | 5134 | 742 | 13.91 |
| 2008 | 369 | 359 | 4252 | 791 | 11.52 |
| 2009 | 453 | 426 | 5182 | 1005 | 11.44 |
| 2010 | 488 | 472 | 6039 | 1015 | 12.38 |
| 2011 | 575 | 552 | 5524 | 1279 | 9.61 |
| 2012 | 854 | 814 | 11034 | 2442 | 12.92 |
| 2013 | 1825 | 1781 | 12328 | 2907 | 6.76 |
| 2014 | 2425 | 2356 | 18225 | 3329 | 7.52 |
| 2015 | 2572 | 2487 | 21740 | 4476 | 8.45 |
| 2016 | 3557 | 3414 | 20644 | 5428 | 5.80 |
| 2017 | 3670 | 3557 | 31566 | 6426 | 8.60 |
| 2018 | 4248 | 4056 | 28648 | 6501 | 6.74 |
| 2019 | 5120 | 4894 | 32874 | 7633 | 6.42 |
| 2020 | 5652 | 5474 | 29109 | 6037 | 5.15 |
| 2021 | 6003 | 5759 | 17685 | 4181 | 2.95 |
| 2022 | 6218 | 6012 | 4912 | 1247 | 0.79 |

where DOC is the total number of documents (publications), CIT (DOC) is the number of cited documents, CIT is the total number of citations, SelfCIT is the number of self-citations and CIT/DOC is the average number of citations per document (publication).

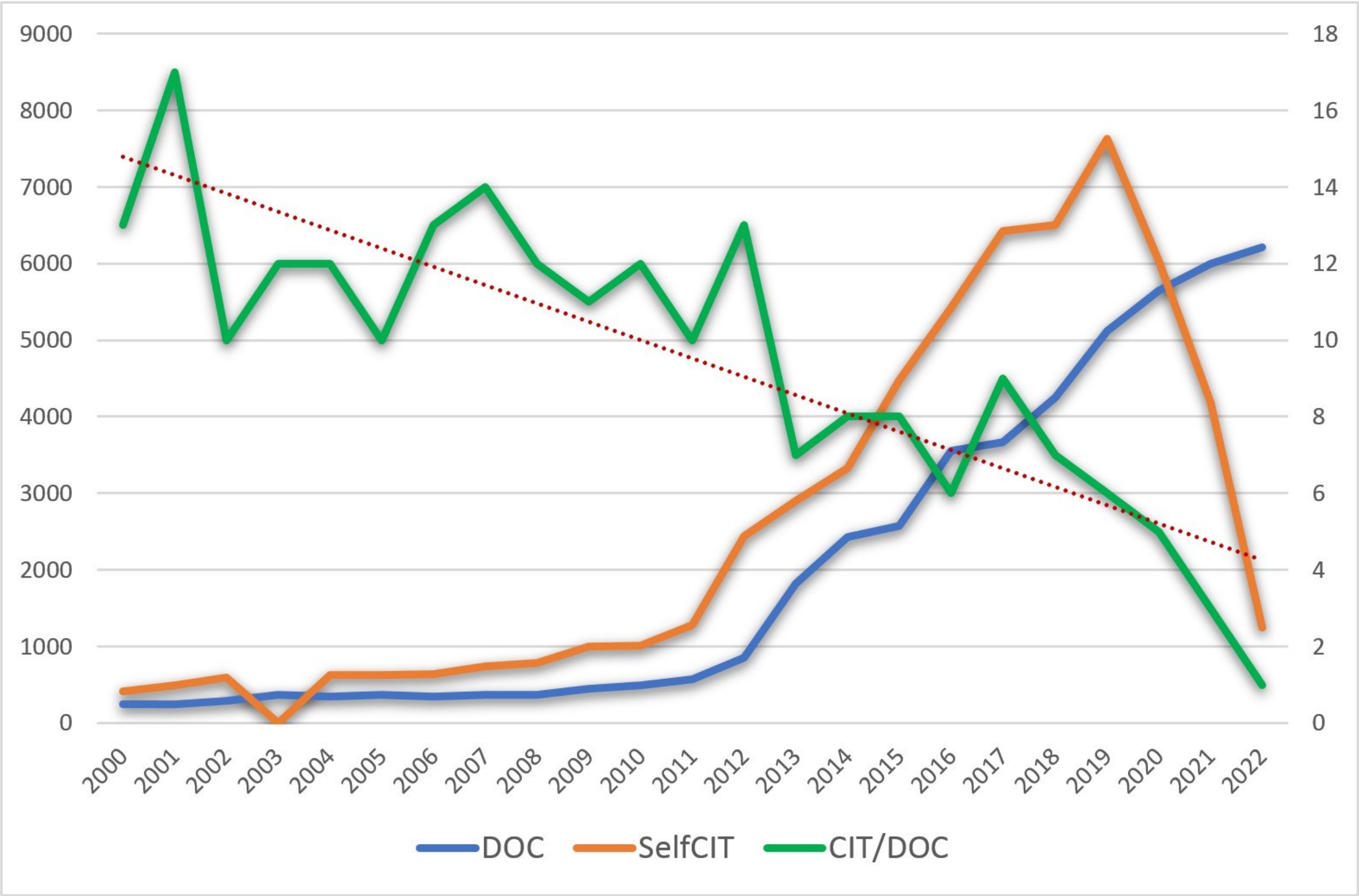

**Fig. 10.** Overall number (in Kazakhstan) of publications, self-citations, and citations per publication by year

With the increasing number of publications, especially stimulated by the Law of the Republic of Kazakhstan dated February 18, 2011, No. 407-IV "On Science," and the introduction of the Hirsch index into circulation, the scientific value of publications, interpreted through citations per document (demand), has systematically declined. This is not only a problem in the science of Kazakhstan but also a problem for all countries where the Hirsch index has dominated as a bibliometric indicator since 2005.

Simple extrapolation shows that maintaining these dynamics and trends will lead to the loss of scientific value for a large portion of publications, and the majority of publications will simply go unnoticed by the scientific community (dark matter of science). Therefore, a different trend is needed: a reduction in the quantity of publications with a corresponding increase in their quality and significance in terms of interpreted cited demand.

The K-index fully aligns with the conditions of this reverse trend by maximizing, according to the Wald criterion (Wald, 1947), the creative role and contribution of coauthors and the individual scientific value of the publication.

Of course, there is a risk and possibility of artificially inflating citations in the K-index. However, this risk and possibility were actualized for the Hirsch index as well. Minimizing and eliminating artificial inflation of citations is possible under the following conditions:

Only citations mentioned in documents indexed in recommended databases (SCOPUS, Web of Science) were considered.

Only citations by documents are considered, not the total sum of citations.

Self-citations are not taken into account.

Citations from close associates (citations from former or current coauthors or from the same institution) are not considered.

Only one citation per author is considered in the source.

Citations of publications of an erroneous, incorrect, or nonscientific nature are not taken into account (this problem can be addressed through artificial intelligence (AI) systems using Popper's criteria algorithms - falsifiability, demarcation, negative character of citation).

The K-index also allows minimizing and neutralizing another danger inherent in the H-index or initiated by the requirements of the Hirsch index.

This involves:

- The necessity of publications in predatory journals to increase the number of articles.
- The abovementioned need for excessive publishing activity is often unjustified.
- The commercialization of scientific publishing activity through "shadow business."

None of these aspects are provided for or minimized in the K-index. Nevertheless, under the extractive conditions of the Kazakhstani scientific system, bureaucratic influence on the K-index is inevitable. "*Power distance*" (Hofstede, 2005) will create conditions where a scientific bureaucracy successfully contends for dominant positions in role functions. In some cases or circumstances, actual contributors to the scientific and creative components are displaced.

Transitioning from the H-index to the K-index will limit the avalanche-like flow of publications, eliminate the need for a large number of publications for the sake of reporting, and focus attention on the scientific value and demand for each individual scientific work.

The Hirsch index may well remain an imperative in international scientific relations. However, within the scientific sovereignty and adjustment of more inclusive stimulating assessments for the Kazakhstani scientific system, a shift to internal dispositional norms in the notation of the K-index is necessary.

Moreover, the H-index may remain on par or in parallel with the K-index. Pilot implementation of the K-index by industry or program is possible. For example, a separate grant program for all scientific fields based on the K-index. Or the transition of grant programs to the K-index for the humanities. Specifically, a philosopher could focus efforts on publishing a single book or monograph systematically presenting new views or paradigms without concern for the possibility of losing "scientific significance" in the context of the Hirsch index.

The K-index can also be interpreted in an expanded form, taking into account the coefficient of international-level scientific insignias and the coefficient of scientific exclusivity (originality), expressed through plagiarism detection systems.

## References

Abramo, G., D'Angelo, C.A., Murgia, G. (2017). The relationship among research productivity, research collaboration, and their determinants. *Journal of Informetrics*, 2017, 11 (4), 1016-1030.

Adams, J. (2013). The fourth age of research. *Nature*, 497, 557–560.

Adams J., J. R. Potter, R., Pendlebury, D., Szomszor, M. (2021). Multiauthorship and Research Analytics/*Handbook for Scientometrics: Science and Technology Development Indicators, Second edition*, pp. 325-346.

Brodie, R. (2004). *Virus of the Mind: The New Science of the Meme.* Integral Pr, p. 251.

Goodhart, C. A. E. (1975). Problems of Monetary Management: The UK experience. *Papers in Monetary Economics,* I, 91-121.

Hirsch, J. E. (2005). An index to quantify an individual's scientific research output. *Proceedings of the National Academy of Sciences of the United States of America*, 102 (46), 16569-16572.

Hofstede, G., Hofstede, G.J., Minkov M. (2005). Cultures and organizations: Software of the mind, revised and expanded. New York: McGraw-Hill, p 576.

Ingham, A.G., Levinger, G., Graves, J., Peckham, V. (1974). The Ringelmann effect: Studies of group size and group performance. *Journal of Experimental Social Psychology*, 10, 371-384.

Lee, S., Bozeman., B. (2005). The Impact of Research Collaboration on Scientific Productivity. *Social Studies of Science,* 35(5), 673–702.

Merton, R. K. (1986). The Matthew Effect in Science. *Science,* 159 (3810), 56-63.

Scarazzati, S., Wang, L. (2019). The effect of collaborations on scientific research output: the case of nanoscience in Chinese regions. *Scientometrics*, 121, 839–868

von Bertalanffy, L. (1962). General System Theory. A Critical Review, «General Systems», VII, pp. 1-20.

von Neumann, J., Morgenstern, O. (1944). *Theory of Games and Economic Behavior.* Princeton University Press, p 776.

Wald, A. (1947). *Sequential Analysis*. New York: John Wiley & Sons, p 212.

Zajonc, R., Murphy, S. (1993). Affect, cognition, and awareness: Affective priming with optimal and suboptimal stimulus exposures. *Journal of Personality and Social Biology*, 64, 723-739.

2007 E. H. Moore Prize (2007). *Notices of the AMS.* p 533